\documentclass[pre,12pt, a4paper, onecolumn, secnumarabic, amssymb, amsfonts, amsmath, bibnotes, superscriptaddress, floatfix, altaffilletter, eqsecnum, final, nofootinbib]{revtex4}
\pdfoutput=1
\usepackage{fullpage}
\usepackage{bm} 
\usepackage{color}
\usepackage{hyperref}
\usepackage{appendix}
\usepackage{graphicx}
\usepackage{enumerate}
\usepackage[english, british]{babel}
\usepackage{float}
\usepackage{eso-pic}
\usepackage{epsfig}
\usepackage[all]{hypcap}
\usepackage{paralist}
\usepackage{subfig}
\usepackage{datetime}
\usepackage{alltt}
\newcommand{\beqs}{\begin{equation}\begin{split}}
\newcommand{\eeqs}{\end{equation}\end{split}}

\begin{document}
\title{SuGra on $G_2$ Structure Backgrounds that Asymptote to $AdS_4$ and Holographic Duals of Confining $2+1d$ Gauge Theories with $\mathcal{N}=1$ SUSY}
\author{Niall T. Macpherson\footnote{\href{mailto:pymacpherson@swansea.ac.uk}{\tt pymacpherson@swansea.ac.uk}}}
\affiliation{Department of Physics, Swansea University\\Singleton Park, Swansea SA2 8PP, United Kingdom}
\begin{abstract}
\vspace{3.0 cm}
\vspace{0.4 cm}
\normalsize{\center{$\bm{Abstract:}$}\vspace{3 mm}\\
In this work the solution generated by performing a U-duality on a deformation of the Maldacena-Nastase solution is studied. This is a solution of type-IIA with a metric that is asymptotically $AdS_4$ and supports a $G_2$ structure. It is believed to be dual to a $2+1d$, $\mathcal{N}=1$ gauge theory similar to the baryonic branch of Klebanov-Strassler with an additional intermediate scale. An improved radial coordinate is used allowing the derivation of UV series solutions to the BPS equation that persist to all orders. A study of the properties of the dual field theory is performed which includes an operator analysis, Wilson loops and a proposal for gauge couplings. The gauge theory dual appears to be a confining Chern-Simons quiver with gauge couplings that become constant at high energies.}\\
\vspace{2.0 cm}
\newpage
\end{abstract}
\vspace{5.0 cm}
\maketitle

\def\tocname{\Large{Table of contents}}


\setlength{\parskip}{1.5mm}
\newpage
\tableofcontents
\cleardoublepage




\newpage

\section{Introduction}

In \cite{Gaillard:2010gy} Gaillard and Martelli introduce a solution generating technique which maps a $\mathcal{N}=1$ type IIA solution supporting $G_2$ a structure into a more general $\mathcal{N}=1$ type IIA solution with interpolating $G_2$ structure. This technique is one of a collection which are generally referred to as "Rotation", as this is how the procedure acts on the space of Killing spinors (see \cite{Gaillard:2010qg,Caceres:2011zn,Elander:2011mh} on $SU(3)$ structure rotations). The rotation of \cite{Gaillard:2010gy} has to be applied to a "seed" solution of type-II SuGra with an unwarped metric, NS 3-form flux and dilaton only. After applying the rotation procedure you are mapped to a type-IIA solution with a warped metric, NS 3-form and RR 4-form flux which will be referred to as the Gaillard-Martelli solution. In \cite{Gaillard:2010gy} the rotation is applied to a deformation, due to Canoura, Merlatti and Ramallo \cite{Canoura:2008at}, of the Maldacena-Nastase solution \cite{Maldacena:2001pb}. They show that the rotation gives rise to a background with NS5 and D2 branes (it is argued in \cite{Macpherson:2012zy} that there are also D4-branes). 

The Maldacena-Nastase solution is dual, in the IR, to the large $N_c$ limit of $2+1$ dimensional $\mathcal{N}=1$ SYM with Chern-Simons level $k=\frac{N_c}{2}$. It is generated by 5-branes which wrap a 3-cycle which grows from zero in the IR to infinity in the UV. Thus in the IR the gauge theory living on the world volume of the branes is 3 dimensional, but as we flow towards the UV the branes unwrap and the world volume gauge theory becomes 6 dimensional. From this it is clear that the Maldacena-Nastase solution is dual to a theory that is not UV complete. The Rotation procedure provides such a completion through extra warp factors introduced into the metric, these ensure that the 3-cycle on which the branes are wrapped remains finite in the UV.

It is important to appreciate the close connection between the Gaillard-Martelli solution and (the baryonic branch \cite{Butti:2004pk} of) Klebanov-Strassler \cite{Klebanov:2000hb}. In \cite{Maldacena:2009mw} it was shown that starting from a deformation \cite{Casero:2006pt}, of the Maldacena-Nunez solution \cite{Maldacena:2000yy}, one can apply an $SU(3)$ rotation and generate the baryonic branch solution. It is U-duality which is discussed in \cite{Maldacena:2009mw} but this is equivalent to rotation, \cite{Gaillard:2010qg}. The Gaillard-Martelli solution is generated from the $2+1$ dimensional equivalent of the deformed Maldacena-Nunez solution and so one expects it to have similar properties. In particular it is probably dual to a cascading 2 node quiver which, being $2+1$ dimensional, will also have a Chern-Simon term. It is interesting to note that whilst the UV of the baryonic branch is Klebanov-Strassler, the UV of the Gaillard-Martelli solution is $AdS_4\times Y$ where $Y$ is the metric at the base of a $G_2$ cone.

In \cite{Gaillard:2010qg} a generalisation of the baryonic branch of Klebanov-Strassler was derived. This was achieved by applying the same $SU(3)$ rotation to the dual of $3+1$ dimensional $\mathcal{N}=1$ SQCD with massless flavours \cite{Casero:2006pt}. It was observed that the numerology of the resulting solution was that of a modified two node quiver with both a duality and a higgsing cascade. There is no separate flavour symmetry (gauged or otherwise) after rotation just a modification to the original gauge groups. However, this solution contains two pathologies, the first is the IR flavour singularity inherited from using \cite{Casero:2006pt} as a seed solution. The second is a rapidly increasing number of D3 branes in the UV that causes the metric to deviate from the desirable Klebanov-Strassler asymptotics. Both these issues are solved in \cite{Conde:2011aa}, where a modified version of the dual of SQCD with massive flavours \cite{Conde:2011rg}, is used as a seed solution. Massive flavours are added in \cite{Conde:2011rg} by means of a flavour profile which interpolates between 0 in the IR, so there are no flavours and thus no singularity, and 1 in the UV where the flavours are effectively massless. The idea of \cite{Conde:2011aa} was to use a profile which also dies off in the UV which allows the inclusion of an additional intermediate scale whilst maintaining the Klebanov-Strassler asymptotics.

With the insight gained from the aforementioned $SU(3)$ structure solutions the dual of $2+1$ dimensional SQCD with massive flavours was derived in \cite{Macpherson:2012zy}. This work also used this as a seed to generate a generalisation of the Gaillard-Martelli solution which is a $G_2$ structure equivalent of \cite{Conde:2011aa}. Here it is necessary to introduce a profile that grows from the IR but dies away again in the UV to maintain the $AdS$ asymptotics. As discussed in \cite{Gaillard:2010gy,Macpherson:2012zy}, there are some technical difficulties in analysing the field theory dual to the Gaillard-Martelli solution and its generalisation. It is not clear how to make the same sort of matching of quiver numerology which is possible for Klebanov-Strassler. This is in part due to it being a running integral of $C_{(3)}$, rather than $B_{(2)}$, which must presumably be used to define the cascade. So it seems that we must use less direct methods to probe the dynamics of the dual gauge theory and this is the focus of this work. 

The main part of the paper is divided into 2, with Section \ref{OnTheSupergravity} concentrating on aspects of the SuGra solution and Section \ref{OntheFieldTheory} focusing on the field theory dual. 

In Section \ref{setup} the generalised Gaillard-Martelli solution is reviewed and an improved radial coordinate is introduced. The BPS equations are then solved in terms of this new radial coordinate in Section \ref{expansions} where, unlike previous attempts, a UV series expansion is derived which persists to all orders. As with the deformed Maldacena-Nunez solution considered in \cite{HoyosBadajoz:2008fw} this expansion is in both exponentials and polynomials. The new radial variable also allows the physically well motivated profile of \cite{Conde:2011aa} to introduce an intermediate scale, and numerical matching of the IR and UV asymptotic solutions is shown for this profile. In Section \ref{ctoinf} it is shown that the metric is asymptotically the product of $AdS_4$ and the compact metric at the base of a $G_2$ cone. Here an exact solution, that can be extracted from a limit of parameter space, is also identified. Section \ref{Charges} then studies the Page and Maxwell charges for the various branes supported by the backgrounds of both the exact and more general solutions.

In the next sections attention is then turned to the field theory dual. In Section \ref{operators} an operator analysis is performed for the first time. Confinement is then shown for the generalised system via a study of Wilson loops in Section \ref{WilsonLoops} were the intermediate scale is also seen via a first order phase transition. Then in Section \ref{couplings} a proposal is made for the couplings of the dual that is consistent with a confining Chern-Simions theory in the IR.

The results are summarised in Section \ref{Diss} were comments on future directions are also made. Finally there are two appendices, Appendix \ref{BPSEQ} giving the BPS equations and Appendix \ref{A2} their general semi analytic UV solution in terms of 4 independent integration constants.

\section{On the Supergravity}\label{OnTheSupergravity}

\subsection{The Type-IIA SuGra Set Up}\label{setup}
The purpose of this section is to briefly review the generalised Gaillard-Martelli solution (more details can be found in \cite{Gaillard:2010gy,Macpherson:2012zy}). The string frame metric is given by:
\begin{equation}\label{met}
ds^2_{\text{str}} = N_c\bigg(\frac{1}{c\sqrt{H}}dx^2_{1,2}+\sqrt{H}ds^2_7\bigg)
\end{equation}
where the internal part of the metric, $ds^2_7$, describes a manifold supporting a $G_2$ structure. This and the warp factor, $H$, are given by:
\begin{equation}\label{Hand7dmet}
\begin{split}
ds^2_7= &  e^{2h}dr^2+\frac{e^{2h}}{4}(\sigma^i)^2+\frac{e^{2g}}{4}(\omega^i-\frac{1}{2}(1+w)\sigma^i)^2\\
H= &1-(\tanh\beta)^2 e^{2(\phi_{\infty}-\phi^{(0)})}
\end{split}
\end{equation}
Notice that we are using a different definition of holographic coordinate, $r$, to that previously used in \cite{Gaillard:2010gy,Macpherson:2012zy}. This will be the key to finding the improved asymptotic series. The functions $g$, $h$, $w$ and $\phi_0$ all depend on $r$ only. The constant $\phi_{\infty}$ is the asymptotic value of $\phi^{(0)}$ in the U.V and $\beta$ parametrises the interpolation of the $G_2$ structure. The constant c is a parameter which enters into the asymptotic UV solutions to the BPS equation (See Eq.\ref{UVexpansions}). $\sigma^i$ and $\omega^i$ are 2 $SU(2)$ left invariant 1-forms which satisfy the following differential relations:
\begin{equation}
d\sigma^i=-\frac{1}{2}\epsilon_{ijk}\sigma^j\wedge\sigma^k;~~~~~d\omega^i=-\frac{1}{2}\epsilon_{ijk}\omega^j\wedge\omega^k
\end{equation}
These can be represented by introducing 3 angles for $\sigma^i$, $(\theta_1,\phi_1,\psi_1)$ and a further 3 for $\omega^i$, $(\theta_2,\phi_2,\psi_2)$ such that:
\begin{equation}\label{rep}
\left.\begin{array}{l l}
\vspace{3 mm}
\sigma^1=&\cos\psi_1 d\theta_1 +\sin\psi_1\sin\theta_1 d\phi_1\\
\vspace{3 mm}
\sigma^2=&-\sin\psi_1 d\theta_1 +\cos\psi_1\sin\theta_1 d\phi_1\\
\vspace{3 mm}
\sigma^3=& d\psi_1+\cos\theta_1 d\phi_1
\end{array}\right.
\end{equation}
and similarly for $\omega^i$. The angles are defined over the ranges: $0\leq\theta_{1,2}\leq\pi$, $0\leq\phi_{1,2}<2\pi$ and $0\leq\psi_{1,2}<4\pi$

This type-IIA solution includes 2 non trivial fluxes, an RR 4-form $F_{(4)}$ and a NS 3-form $H_{(3)}$. They are given by:
\begin{equation}\label{H3F4}
\left.\begin{array}{l l}
\vspace{3 mm}
F_{(4)}=&\!\!- N_c^{3/2}e^{-\phi_{\infty}}\frac{\sqrt{\cosh(\beta)}}{c^{3/2}\sinh(\beta)} Vol_{(3)}\wedge dH^{-1}+ N_c^{1/2}\frac{\tanh(\beta)}{\sqrt{\cosh(\beta)}}e^{\phi_{\infty}-2\phi^{(0)}}*_7 H_{(3)}\\
H_{(3)}=&\frac{N_c}{4}\bigg[\left((\kappa+\frac{1}{2}\!+\!\frac{3x}{2}(C\!-\!1)P)\sigma^1\wedge\sigma^2\wedge\sigma^3-\omega^1\wedge\omega^2\wedge\omega^3\right)\!+\\
&~~~~~~\frac{4xP'\eta+\gamma'}{2}dr\wedge\sigma^i\wedge\omega^i +\\&
~~~~~~\frac{1}{4}\epsilon_{ijk}\left(\left(1\!+\!\gamma\right)\sigma^i\wedge\sigma^j\wedge\omega^k\!-\!\left(1\!+\!\gamma\!-\!2xP\right)\omega^i\wedge\omega^j\wedge\sigma^k\right)\bigg]
\end{array}\right.
\end{equation}
Where the function $\gamma$ depends on $r$ only, and $\eta$ is given by Eq.\ref{VAndeta}. $\kappa$ and $C$ are constants that will be fixed below. The dilaton is given by:
\begin{equation}
e^{2\phi}=\cosh(\beta)e^{2\phi^{(0)}} H^{1/2} 
\end{equation}

The function P is a profile which generalises the solution considered in \cite{Gaillard:2010gy}. In this work it interpolates between zero in the IR and grows to a maxima before tending smoothly back to zero in the UV. This introduces an intermediate scale into the theory that is parametrised by the constant $x$. From the smeared flavour brane origin of this solution (see \cite{Macpherson:2012zy}) it is natural to set $x=\frac{N_f}{N_c}$ but as we do not expect the dual field theory to have an explicit flavour symmetry\footnote{See also \cite{Conde:2011aa} in the context of the conifold.} this point will not be laboured. A smeared brane configuration is introduced via a violation of the Bianchi identity for $H_{(3)}$:
\begin{equation}
\left.\begin{array}{l l}
dH_{(3)}&=\Xi_4=\\
&~~~~-\frac{xN_c}{4}\bigg[\frac{1}{4}\epsilon_{ilm}\epsilon_{ijk}\sigma^l\wedge\sigma^m\wedge\omega^j\wedge\omega^k+ \eta P'\epsilon_{ijk}dr\wedge\sigma^i\wedge\omega^j\wedge\omega^k\\
&~~~~-\frac{3}{2}(C-1)P'dr\wedge\sigma^1\wedge\sigma^2\wedge\sigma^3-\frac{1}{2}(2\eta+1)P'\epsilon_{ijk} dr\wedge\sigma^i\wedge\sigma^j\wedge\omega^k\bigg]\\
\end{array}\right.
\end{equation}

When $P=0$ the solution considered in \cite{Gaillard:2010gy} is recovered. $P=1$ is the solution generated by applying the rotation procedure to the massless flavour solution derived in \cite{Canoura:2008at}. This has a singularity in the IR and a fast growing number of D2 branes in the UV which gives undesirable asymptotics\footnote{This was first observed in \cite{Gaillard:2010qg} for D3 branes on the conifold}.
In \cite{Macpherson:2012zy}, these issues are resolved by a profile which kills off the $H_{(3)}$ source in both the far IR and UV.

From this point on:
\begin{equation}
C=1; ~~~~ \kappa=\frac{1}{2}
\end{equation}
C is set to this value to enable the dual gauge theory to have a quantised Chern-Simons term \cite{Macpherson:2012zy}. $\kappa$ must be thus set so as to avoid a curvature singularity in the IR \cite{Maldacena:2001pb,Macpherson:2012zy}.

This work will focus on the limit $\beta\to\infty$ which will require that the following identification is made:
\begin{equation}\label{srelation}
c = e^{2\phi_{\infty}}\sinh\beta 
\end{equation}
As explained in \cite{Gaillard:2010gy} this combination has to be held fixed if the limit $\beta\to\infty$ is to be well defined\footnote{It should be noted however that in  \cite{Gaillard:2010gy} it was a different constant that was held fixed. What was referred to as c there is $g_0$ here in \ref{IRExpansion1}}. This allows the factors of $\cosh(\beta)$ and $\sinh(\beta)$ in the dilaton and RR 4-form to be replaced with $\tanh(\beta)$, $e^{\phi_{\infty}}$ and $c$ such that they remain finite when $\beta\to\infty$. After taking the limit:
\begin{equation}
\left.\begin{array}{l l}
\vspace{3 mm}
e^{2\phi}=&c H^{1/2} e^{2(\phi^{(0)}-\phi_{\infty})}\\

F_{(4)}=&-\frac{1}{c^2} N_c^{3/2} Vol_{(3)}\wedge dH^{-1}+\frac{1}{c^{1/2}} N_c^{1/2}e^{2(\phi_{\infty}-\phi)}*_7 H_{(3)}
\end{array}\right.
\end{equation}
Specifically the independent components of $F_{(4)}$ are given by:
\begin{equation}\label{equation:ComponentsofF4}
\left.
  \begin{array}{l l }
    \vspace{3 mm}
F^{(4)}_{r123} = -\frac{2}{\sqrt{c N_c}H}e^{-3g+2(\phi_\infty-\phi^{(0)})};&F^{(4)}_{r\hat{i}\hat{j}k} = \frac{1}{2\sqrt{c N_c}H}\epsilon_{ijk}\left(1+w^2-4xP -2w\gamma\right)e^{-g-2h+2(\phi_\infty-\phi^{(0)})};\\
  \vspace{3 mm}
F^{(4)}_{r\hat{1}\hat{2}\hat{3}} = -\frac{1}{\sqrt{4c N_c} H} V e^{-3h+2(\phi_\infty-\phi^{(0)})};&F^{(4)}_{r\hat{i}jk} =\frac{1}{\sqrt{c N_c}H}\epsilon_{ijk}\left(w-\gamma\right)e^{-2g-h+2(\phi_\infty-\phi^{(0)})};\\
  \vspace{3 mm}
F^{(4)}_{txyr} = -\frac{H'}{\sqrt{c N_c} H^{3/2}};
&F^{(4)}_{i\hat{i}j\hat{j}} = \frac{1}{2 \sqrt{c N_c}H}\left(4x \eta P'+\gamma'\right)e^{-2g-h+2(\phi_\infty-\phi^{(0)})};\\
\end{array}\right.
\end{equation}
where $V$ and $\eta$ are defined in Eq.\ref{VAndeta}. The components are expressed in the following vielbein basis:
\begin{equation}
\left.\begin{array}{l l}
e^{x^i}=\sqrt{c N_c}H^{-1/4}dx^i,~~~~~~& e^r = \sqrt{N_c}H^{1/4}e^{g} dr,\\
e^i= \sqrt{N_c} H^{1/4} e^{h}\frac{\sigma^i}{2},~~~~~& e^{\hat{i}} = \sqrt{N_c}H^{1/4}e^{g}\left(\frac{\omega^i - \frac{1}{2}(1+w)\sigma^i}{2}\right)
\end{array}\right.
\end{equation}
Finally, a potential $C_{(3)}$ such that $dC_{(3)}=F_{(4)}$ is given by:
\begin{equation}\label{equation:ComponentsofC3}
\left.
  \begin{array}{l l }
    \vspace{3 mm}
C_{123}^{(3)}=\frac{ e^{2(\phi_\infty-\phi^{(0)})}}{\sqrt{c}H^{3/4}}\cos\alpha;&~~~C_{i\hat{j}\hat{k}}^{(3)}=\epsilon_{i\hat{j}\hat{k}}\frac{ e^{2(\phi_\infty-\phi^{(0)})}}{\sqrt{c}H^{3/4}}\cos\alpha\\
 \vspace{3 mm}
 C_{\hat{1}\hat{2}\hat{3}}^{(3)}=\frac{ e^{2(\phi_\infty-\phi^{(0)})}}{\sqrt{c}H^{3/4}}\sin\alpha;&~~~C_{i\hat{j}\hat{k}}^{(3)}=-\epsilon_{i\hat{j}\hat{k}}\frac{ e^{2(\phi_\infty-\phi^{(0)})}}{\sqrt{c}H^{3/4}}\sin\alpha\\
  \vspace{3 mm}
  C_{txy}^{(3)}=\frac{1}{\sqrt{c}H^{1/4}};&~~~C_{r i\hat{i}}^{(3)}=\frac{ e^{2(\phi_\infty-\phi^{(0)})}}{\sqrt{c}H^{3/4}}\\
\end{array}\right.
\end{equation}
Where $\alpha$ is defined in Eq.\ref{alpha}.

\subsection{Solutions to the BPS Equations}\label{expansions}

In this section we present improved (with respect to \cite{Canoura:2008at,Gaillard:2010gy,Macpherson:2012zy}) solutions to the BPS equations of Appendix \ref{BPSEQ} that give rise to asymptotically constant $\phi^{(0)}$. The claim is not that these are new solutions but rather they are expressed in terms of the optimum holographic variable.

It is only possible to solve the BPS equations of Appendix \ref{BPSEQ} analytically in terms of series expansions in the IR and the UV and then to show that these series can be smoothly connected over the full range of the holographic coordinate numerically with a shooting method. This was done for the profile dependent solution considered here in \cite{Macpherson:2012zy}, but in terms of a different holographic coordinate\footnote{Specifically $\rho=e^{2h}$, were in Eq.\ref{met} $e^h dr_{\text{here}}= dr_{\text{there}}$} (see also \cite{Gaillard:2010gy,Canoura:2008at} for $P=1$ and $P=0$ cases respectively). 

In previous works the UV series solution were only found to the first few orders in terms of polynomials and did not depend on a UV integration constant, while the IR did. This meant that one had to shoot from the IR quite far into the UV in order to match the asymptotic series solutions. It is possible to show that the full IR expansion in terms of the holographic coordinates of \cite{Gaillard:2010gy,Canoura:2008at,Macpherson:2012zy} is in terms of polynomials and logarithms. However the choice of $r$ in Eq.\ref{met} gives a semi-analytic UV solution in terms of exponentials and polynomials. This is far easier to work with as the numerical and series expansions converge much faster and there are independent tunable parameters in both the IR and UV. The relation between $G_2$-cone and conifold solution is also some what illuminated as they have UV solutions of the same form.

Although asymptotic solutions can be derived for quite general profile functions the specific choice of \cite{Macpherson:2012zy} is no longer valid\footnote{It does not have an expansion in exponentials as $r\to\infty$}. The obvious choice is now the profile of \cite{Conde:2011aa}:
\begin{equation}\label{eq:pro}
P(r)=\theta(r-r_*)\tanh^42(r-r_*)e^{-4/3(r-r_*)}
\end{equation}
This is well motivated physically for reasons that are explained at length in \cite{Barranco:2011vt,Conde:2011aa}\footnote{Modulo small differences as these works refer to D5-brane sources on the conifold.}, the basic argument is repeated below.

When one expresses the deformed Maldacena-Nastase in terms of the present radial variable, as shown in Eq.\ref{UVexpansions}, one is lead to a UV series expansion in terms of exponentials and polynomials when the dilaton is UV constant. This solution has metric and forms given by Eqs.\ref{met}\ref{H3F4} with $\beta=0$ which sends $H\to1$, so the is no longer a warp factors on the metric and $F_{(4)}=0$. In order for massive flavours to be added, and so as to avoid a flavour singularity in the IR, one would require a profile such that $P \sim r^4$ in the IR and $P \sim 1$ in the UV. This ensures that the solution will be deformed Maldacena-Nastase in the IR and a massless flavour solution in the UV. A simple choice is $P(r)=\theta(r-r_*)\tanh^42(r-r_*)$ where the maximum of $P'$ gives a measure of the common quark mass\footnote{Or it would if there was an ungauged flavour symmetry. When $H=1$ this is only the true for an asymptotically linear dilaton, as it is only for this case that putative flavour states are normalisable.} . 

If one now applies the rotation algorithm of \cite{Gaillard:2010gy} to this and generate a generalisation of the Gaillard-Martelli solution a problem arises. As shown in \cite{Macpherson:2012zy}, the rotation procedure induces D2 branes charge on the NS5 branes. When $P\sim1$ in the UV the D2 brane charge builds up very quickly and causes a back reaction on the geometry which deforms the solution away from $AdS_4$. A similar effect is also seen in \cite{Gaillard:2010qg} where it is a build up of D3 charge on D5 branes. In the field theory dual this is equivalent to inserting an irrelevant operator into the Lagrangian. $AdS$ asymptotics are highly desirable as they allow one to perform holographic renormalisation group techniques, and so this needs to be rectified. To achieve this one requires $P\sim0$ in the UV, and for phenomenological reasons the choice in Eq.\ref{eq:pro} is very attractive. This is the minimal choice that restores the $AdS_4$ asymptotics and keeps an intermediate scale governed by $P$. Greater suppression would also achieve $P\sim0$ but can generically change the sub leading form of the UV solutions. This would turn on additional vevs in the dual field theory in a fashion that is hard to justify physically. Greater suppression was also shown to lead to undesirable effects, such as a non monotonic central charge, in the similar conifold solution of \cite{Conde:2011aa}.

With a profile chosen all that remains is to present the solutions to the BPS equations. In what follows $r_*$ is a constant which can be varied and defines where the new scale enters into the theory. The transition between $r<r_*$ and $r>r*$ is continuous and so there is no singularity at $r=r_*$, this is because $P$, $P'$, $P''$, $P'''$ match as $r \to r_* +$ and $r \to r_* -$.

When $r_*=0$ the new scale begins to appear at $r=0$ and the IR expansion is given by: 
\begin{equation}\label{IRExpansion1}
\left.\begin{array}{l l}      

\vspace{3 mm}             
e^{2g}&=g_0+\frac{(g_0-1) (9 g_0+5)
   }{12 g_0}r^2+\bigg[\frac{(g_0-1)
   \left(54 g_0^3+30 g_0^2+25 g_0+29\right)}{432
   g_0^3}-\frac{8 \left(6 g_0^2-3 g_0+1\right)
   x}{3 g_0^2}\bigg]r^4+\\
\vspace{3 mm}   
   &~~~~\frac{8 \left(203 g_0^2-100 g_0+41\right) 
   x}{105 g_0^2}r^5+ ...\\
\vspace{3 mm}

e^{2h}&=g_0 r^2-\left(\frac{3 g_0^2-4 g_0+4}{18 g_0}+32 x\right) r^4+32x r^5+...\\

\vspace{3 mm}                 
w&=1-\frac{3 g_0-2 }{3 g_0}r^2+ \bigg[\frac{72 g_0^3-84 g_0^2-17 g_0+38}{108
   g_0^3}+\frac{16 (2 g_0-1) x}{g_0^2}\bigg]r^4-\frac{32 (21 g_0-10)  x}{21 g_0^2}r^5+...\\

\vspace{3 mm}
\gamma&=1-\frac{r^2}{3}+\bigg[\frac{9 g_0^2+4 g_0-4}{108 g_0^2}+\frac{16 (3 g_0-2) x}{3 g_0}\bigg]r^4-\frac{32 (14 g_0-11)  x}{21
   g_0}r^5+...\\         
\vspace{3 mm}  
\phi^{(0)}&=\phi_0 -\frac{7+576 g_0 x}{24 g_0^2}r^2+\frac{64  x}{3 g_0}r^3+\bigg[\frac{\left(210 g_0^2+56 g_0-223\right) g_0^4}{1728}+\frac{2
   \left(71 g_0^2-1\right) g_0^5 x}{3}\bigg]r^4+ ...\\      
\end{array}\right.
\end{equation}
The IR expansion when $r_*>0$ is given by Eq.\ref{IRExpansion1} with $x=0$. A series expansion of this type appears to persist to all orders in $r$.

The expansion in the UV takes the following form:
\begin{equation}\label{UVexpansions}
\left.\begin{array}{l l} 
\vspace{3 mm}             
e^{2g}&= ce^{4 r/3}-1 +\frac{33 }{4 c}e^{-4 r/3} -\frac{3168-392 x c}{72 c^2}e^{-8 r/3}+\\
\vspace{3 mm}       
&~~~~\bigg[\frac{-840 c^2 c_{\gamma }-35840
   r+1860768}{7200 c^3}+\frac{ \left(13440 r c-392808 c\right)x}{7200 c^3}+\frac{33 x^2}{40 c}\bigg]e^{-4r}+...\\
\vspace{3 mm}             
e^{2h}&=\frac{3}{4}c e^{4 r/3} +\frac{9}{4}- \left(\frac{77}{16 c}-3 x\right)e^{-4 r/3}+\frac{1536-88 x c}{96
   c^2}e^{-8r/3}+\\
   \vspace{3 mm}
  &~~~~~ \bigg[\frac{360 c^2 c_{\gamma }+15360 r-398912}{3200
   c^3}+\frac{x \left(168072 c-5760 r c\right)}{3200 c^3}-\frac{1053 x^2}{160 c}\bigg]e^{-4r}+...\\
   \vspace{3 mm}
w&=\frac{2
   }{c}e^{-4 r/3}+\frac{ 22-6 x c}{c^2}e^{-8 r/3}+\frac{51-16 x c}{2 c^3}e^{-4 r}+...\\
\vspace{3 mm}
\gamma&=\frac{1}{3}+xe^{-4 r/3}+\bigg[\frac{ 3 c^2 c_{\gamma }+128 r}{3 c^2}-\frac{16 r x}{c}\bigg]e^{-8 r/3}
   +\\
    \vspace{3 mm}
   &~~~~\bigg[\frac{96-6 c^2 c_{\gamma }-256 r}{3 c^3}+\frac{ \left(96 r c-28 c\right)}{3 c^3}x\bigg]e^{-4r}+...\\         
\vspace{3 mm}  
\phi^{(0)}&= \phi _{\infty }+ \frac{8+6 x c}{4 c^2}e^{-8 r/3}-\frac{2 \left(2+x c\right)}{c^3}e^{-4 r}+...\\     
\end{array}\right.
\end{equation}
A series of this type appears to persist to all orders in $r$. In fact there is a more general series for which this is true presented in Appendix \ref{A2}. 

Notice that like the deformed Maldacena-Nunez solution considered in \cite{HoyosBadajoz:2008fw} these expansions are in both exponentials and polynomials. However the polynomial terms start at a more suppressed order. It is for this reason that the metric of this solution is asymptotically AdS rather than Log corrected AdS like (the Baryonic Branch of) Klebanov-Strassler \cite{Klebanov:2000hb,Butti:2004pk}, more on this in the next section.

There is in fact another UV expansion that solves the BPS equations. This gives rise to an asymptotically linear dilaton which, being unbounded, violates the reality of the metric\footnote{Notice in Eq.\ref{Hand7dmet} that if the dilaton is unbounded then $H$ can become negative. When this happens the $H^{1/2}$ factors in Eq.\ref{met} become imaginary} and so we do not consider it here (See \cite{Maldacena:2001pb,Canoura:2008at,Macpherson:2012zy} for details of solutions of this type)

Using the  UV series expansions in Eq.\ref{UVexpansions} it is possible to derive the asymptotic behaviours of both the dilaton $\phi$ and the warp factor $H$ which will be useful in the next section. They are given by:
\begin{equation}\label{hphiUV}
\left.\begin{array}{l l} 
    \vspace{3 mm}
e^{2\phi}&=  \sqrt{3 c
   x+4}e^{-4 r/3}-\frac{2   (c
   x+2)}{c\sqrt{3 c x+4}}e^{-8 r/3}+\frac{5696+9876 c x+4467 c^2 x^2+216 c^3 x^3}{48
  c^2 (3 c x+4)^{3/2}}+...\\
H&=\frac{ 3c x+4}{c^2}e^{-8 r/3}-\frac{4 (c x+2)}{c^3}e^{-4
   r}+\frac{ 3 c x (91-120 c x)+752}{24 c^4}e^{-16 r/3}+...
   \end{array}\right.
\end{equation}

\begin{figure}
 \begin{center}
\includegraphics[width=1\textwidth]{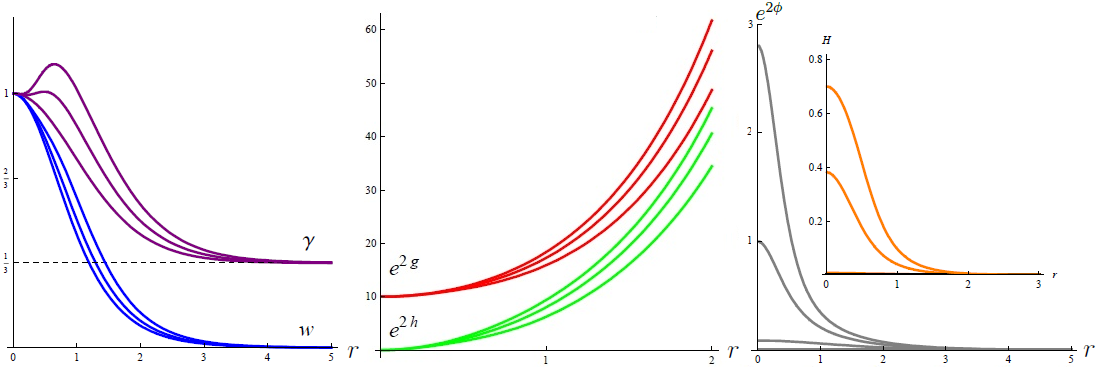} \label{BPSNumerics}
\caption{The graphs are numerical plots of the various functions of the solution for $g_0 = 10$, $x=0,1/2,1$, $r_*=0$. Increasing $g_0$ and x have different effects on each function and they have been grouped together by there behaviour. In the left panel the direction of increasing x is up. The IR and UV behaviours are independent of $g_0$. The width of the distribution of $w$ solutions decreases as $g_0$ increases but $\gamma$ does not change noticeably. Down is the direction of increasing x in the middle plot. The distributions of both $e^{2g}$ and $e^{2h}$ decrease as $g_0$ increases and the UV behaviour depends on $g_0$ as well. Increasing x is once more down in the right panel. As $g_0$ increases the IR value of $H$ and $e^{2\phi}$ decreases as does the width of the distribution of each function. The UV values do not change.}
 \end{center}
\end{figure}

In Fig.\ref{BPSNumerics} there are plots of the numerical solutions of the BPS equations, dilaton and H. These confirm that it is possible to smoothly connect the IR series expansions to the UV series expansion and also show the effect of varying $x$. It is worth making a comment about how these solutions (and others) can be generated. A class of solutions can be defined by choosing values for $x$ and $r_*$ which parametrise the size of the intermediate scale and where it begins. The idea is to use Eq.\ref{IRExpansion1} to define IR boundary conditions for the BPS system at $r_{min}$ close to zero then use a shooting technique to match the numerical solutions to Eq.\ref{UVexpansions} for some range below a large but finite $r_{max}$. Given $(x,r_*)$ there will be a minimum value $g_0=g_{min}$ below which all the numerical solutions of the BPS equations become badly singular and cannot be connected to a sensible UV expansion. When $g_0=g_{min}$ we are led to asymptotically linear behaviour of $\phi^{(0)}$ mentioned above. For all $g_0>g_{min}$ we are led to a valid UV of the type in Eq.\ref{UVexpansions} where $c$ must be tuned to match each particular $(g_0,x,r_*)$ combination. For all the solutions considered here it turned out that $c_\gamma =0$. Thus given $(x,r_*)$ there appears to be exactly one tunable parameter in each of the IR and UV where $c$ increases non linearly with $g_0$.

\subsection{$AdS_4$ Asymptotics and an Exact Solution for Infinite c}\label{ctoinf}
Using the UV series solutions of the previous section it is possible to show that the metric tends to the following form in the UV:
\begin{equation}\label{UVmet}
ds^2 =\frac{9\sqrt{4+3c x}N_c}{4}\bigg[\frac{4e^{4r/3} dx_{1,2}^2}{9(4+3c x)}+\frac{4}{9}dr^2+\frac{1}{12}(\sigma^i)^2 +\frac{1}{9}(\omega^i-\sigma^i/2)^2\bigg]+O(e^{-4r/3}/c)
\end{equation}
which is actually $AdS_4$ in disguise with an additional compact piece which describes the metric at the base of a $G_2$ cone. This can be elucidated with the following rescaling and coordinate transformations:
\begin{equation}\label{redefs}
x^{\mu}\to \frac{3\sqrt{4+3c x}}{2}x^{\mu};~~~~~~~~\rho=e^{2/3r}
\end{equation}
In \cite{Gaillard:2010gy} the $P=0$ case of this solution is studied with a different holographic coordinate. They show that when a certain parameter is taken to infinity an exact solution can be extracted that is $AdS_4\times Y$, where $Y\sim S^3\times S^3$. This solution describes the UV of Gaillard-Martelli but the IR is clearly very different. This exact solution is not conformal as the dilaton depends on their holographic coordinate $r'$ as $\phi =-\frac{1}{2} \log(\frac{2r'}{9})$ which means it actually diverges at $r'=0$. Its is possible to do something similar here and the new radial coordinate means that the dilaton of the exact solution is finite at $r=0$ (See Eq.\ref{exsol}). As discussed in \cite{Gaillard:2010gy,Macpherson:2012zy} there are some complications in interpreting the field theory dual to the Gaillard-Martelli solution and its generalisation and so it seems sensible to attempt to gain some insight from the taking a similar limit here.

The relevant limit is $c\to\infty$, as when this is taken the UV metric in Eq.\ref{UVmet} becomes exact and finite. By construction the profile tends to zero in the UV and although x enters into the Eq.\ref{UVmet} this is only as an overall factor which will not qualitatively change the physics. Thus it is $c\to\infty$ with $x=0$ that will be explored. The metric of the exact solution is given by:
\begin{equation}\label{UVmet2}
ds_{ex}^2 =\frac{9N_c}{2}\bigg[\frac{1}{9} dx_{1,2}^2e^{4r/3}+\frac{4}{9}dr^2+\frac{1}{12}(\sigma^i)^2 +\frac{1}{9}(\omega^i-\sigma^i/2)^2\bigg]
\end{equation}
While the dilaton, RR 4-form and NS 3-form are given by:
\begin{equation}\label{exsol}
\left.\begin{array}{l l} 
    \vspace{3 mm}
    \phi^{ex}&=\frac{1}{2}\bigg[\log(2)-4r/3\bigg]\\
        \vspace{3 mm}
    H^{ex}_{(3)}&=\frac{N_c}{4}\bigg[\sigma^1\wedge\sigma^2\wedge\sigma^3-\omega^1\wedge\omega^2\wedge\omega^3\!+
\frac{1}{4}\epsilon_{ijk}\left(\sigma^i\wedge\sigma^j\wedge\omega^k\!-\!\omega^i\wedge\omega^j\wedge\sigma^k\right)\bigg]\\
        \vspace{3 mm}
        F^{ex}_{(4)}&=-\frac{2N_c^{3/2}}{3}e^{8r/3}Vol_{(3)}\wedge dr-\frac{N_c^{3/2}e^{2r/3}}{4\sqrt{3}}dr\wedge\bigg[\sigma^1\wedge\sigma^2\wedge\sigma^3+\omega^1\wedge\omega^2\wedge\omega^3\bigg]\\
   \end{array}\right.
\end{equation}
A minimal choice for a potential such that $dC_{(3)}=F_{(4)}$ is given by:
\begin{equation}\label{C3ex}
C^{ex}_{(3)}=\frac{N_c^{3/2}e^{8r/3}}{4}Vol_{(3)}-\frac{\sqrt{3}N_c^{3/2}e^{2r/3}}{8}\bigg[\sigma^1\wedge\sigma^2\wedge\sigma^3+\omega^1\wedge\omega^2\wedge\omega^3\bigg]\\
\end{equation}

\subsection{Page and Maxwell Charges}\label{Charges}
The purpose of this section is to learn something about the charges of the respective branes in this type-IIA solution. (see \cite{Marolf:2000cb} for a discussion on Page and Maxwell charges) As pointed out in \cite{Macpherson:2012zy} we have an $F_{(4)}$ which is both an electric and magnetic brane source as well as $H_{(3)}$. So the branes in this solution are $NS5$, $D2$ and $D4$ branes. In \cite{Macpherson:2012zy} it was shown that the Maxwell charge is running for the $D2$ and $D4$ branes and that it was only possible to define a page charge for the $D2$ branes. Thus we will attempt to see what can be learnt from the exact solution of Section \ref{ctoinf} before proceeding any further.

It is clear from \ref{exsol} that for the exact solution there is no cycle on which to define a D4 brane charge and so it must be zero. If we integrate the flux of $H^{ex}_{(3)}$ over the cycle $\omega^1\wedge\omega^2\wedge\omega^3$ we get:
\begin{equation}
-\frac{1}{4\pi^2}\int H^{ex}_{(3)}=N_c
\end{equation}
which gives a quantised NS5 brane charge. The Maxwell charge of the D2 brane will clearly run as the pull back of $*F_{(4)}$ onto the only suitable cycle on which to define this charge is:
\begin{equation}
-\frac{\sqrt{3}}{16}N_c^{5/2}e^{2r/3}\sigma^1\wedge\sigma^2\wedge\sigma^3\wedge\omega^1\wedge\omega^2\wedge\omega^3
\end{equation}
However this is precisely equal to the pull back of $H^{ex}_{(3)}\wedge C^{ex}_{(3)}$ onto the same cycle\footnote{This was chosen rather than $B_{(2)}\wedge F_{(4)}$ for two reasons. As observed in \cite{Gaillard:2010gy} for the solutions considered here $C_{(3)}$ appears to pay something like the role $B_{(2)}$ does in Klebanov-Strassler. But also when we add the source for $H_{(3)}$, $B_{(2)}$ can no longer be defined.}. So it is possible to define a page charge for the D2-branes:
\begin{equation}\label{Mcex}
M^{ex}_c=\frac{1}{(2\pi)^5}\int(*F^{ex}_{(4)}-H^{ex}_{(3)}\wedge C^{ex}_{(3)})=0
\end{equation}
This however is not really the whole story because $C^{ex}_{(3)}$ is not gauge invariant. So large gauge transformations can induce quantised shifts in $M^{ex}_c$. Consider such a large gauge transformation, $C^{ex}_{(3)}\to C^{ex}_{(3)}+\Delta C^{ex}_{(3)}$, where:
\begin{equation}\label{gauge}
\Delta C^{ex}_{(3)}=-\frac{n\pi}{4}\bigg[\sigma^1\wedge\sigma^2\wedge\sigma^3+\omega^1\wedge\omega^2\wedge\omega^3\bigg]
\end{equation}
which is the 3-form equivalent of the gauge transformation considered in \cite{Benini:2007gx}. Under such a gauge transformation we have
\begin{equation}
M^{ex}_c\to M^{ex}_c+ N_c
\end{equation}
with $N_c$ remaining invariant. This is rather reminiscent of the story in Klebanov-Strassler except $C_{(3)}$ is playing the role $B_{(2)}$ usually would. Indeed from Eq.\ref{exsol} it is clear that there is no cycle on which the flux of $B_{(2)}$ runs. However the flux of $C^{ex}_{(3)}$ as written in Eq.\ref{C3ex} runs on the cycle $\sigma^{i}=\omega^{i}$:
\begin{equation}
c^{ex}_0=-\frac{1}{4\pi^2}\int C^{ex}_{(3)}=\sqrt{3}N_c^{3/2}e^{2r/3}
\end{equation}
and of course large gauge transformations will shift this as $c^{ex}_0\to c^{ex}_0 +2n\pi$.

Now consider the full solution of Section \ref{setup} the NS5 brane charge remains unchanged. A D2 brane Page charge can once more be defined by integrating $*F_{(4)}-H_{(3)}\wedge C_{(3)}$ over the same cycle as Eq. \ref{Mcex}. Further more, performing the same large gauge transformation on $C_{(3)}$, Eq. \ref{gauge}, shifts this page charge by the same quantised amount, $M_c\to M_c+ n N_c$. 

\begin{figure}
 \begin{center}             
\includegraphics[width=1\textwidth]{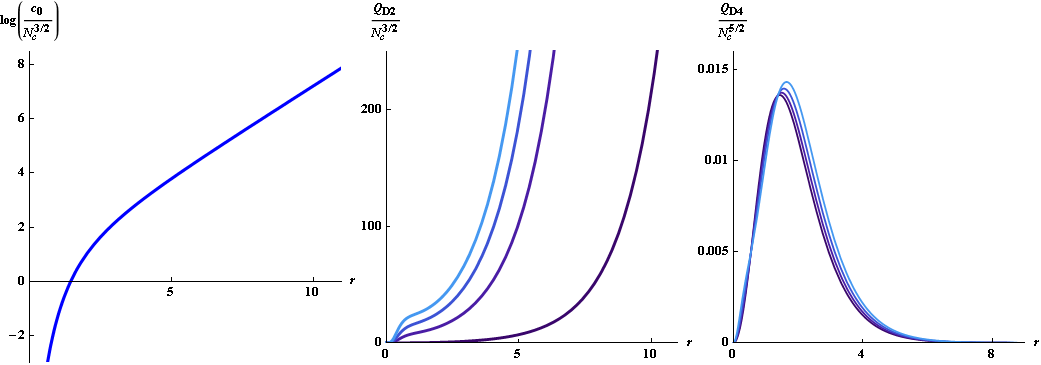}    
  \caption{Plots of SuGra observables. The left panel is plots of $Log(c_0)$ for $x=0$, $r_*=0$ with $g_0=30$. The middle and right panels contains plots of the Maxwell charges $Q_{D2}$ and $Q_{D4}$ for $g_0=40$, $r_*=0$ and $x=0,1,2,3$ colour coded purple to blue.} 
  \label{charges}
 \end{center}
\end{figure}

The flux of $C_{(3)}$ on $\sigma^i=\omega^i$ is no longer exact but is still running. Its asymptotic values are:
\begin{equation}\label{c0}
c_0=-\frac{1}{4\pi^2}\int C_{(3)}= \left\{\begin{array}{l l}
\vspace{1 mm}
N_c^{3/2}e^{2\Delta\phi}\bigg[\frac{g_0}{2\sqrt{c}}r^3+\frac{4 \left(4-3 g_0\right) g_0-1}{24 \sqrt{c g_0}}r^5+...\bigg]&\quad r\approx0\\
\vspace{1 mm}
 \sqrt{3}N_c^{3/2} e^{2 r/3}\bigg[1+\frac{1 }{2 c}e^{-4 r/3}+...\bigg]&\quad r\approx\infty\\
\end{array}\right.
\end{equation}
where $\Delta\phi=\phi_{\infty}-\phi_0$. Similarly we can define a D2 brane Maxwell charge which runs on: $\sigma^1\wedge\sigma^2\wedge\sigma^3\wedge\omega^1\wedge\omega^2\wedge\omega^3$:
\begin{equation}\label{Q2}
Q_{D2}=\frac{1}{(2\pi)^5}\int *F_{(4)}= \left\{\begin{array}{l l}
\vspace{1 mm}
\frac{N_c^{5/2}e^{2\Delta\phi}}{\sqrt{c}\pi}\bigg[\frac{ \sqrt{g_0}  \left(576 g_0 x+7\right)}{48}r^4-16 g_0^{3/2} x r^5+...\bigg]&\quad r\approx0\\
\vspace{1 mm}
\frac{\sqrt{3}N_c^{5/2}}{\pi}\bigg[\frac{3 c x+4}{8}e^{2 r/3}+\frac{9 c x+4}{16 c}e^{-2 r/3}+...\bigg]&\quad r\approx\infty\\
\end{array}\right.
\end{equation}
Numerical plots of $c_0$ and $Q_{D2}$ can be seen in Fig.\ref{charges}. Note that since these quantities contain the UV parameter $c$, this must be matched to $(g_0,x,r_*)$ for each specific solution as described towards the end of Section \ref{expansions}.

For the general solution it is also possible to define a running D4 brane Maxwell charge:
\begin{equation}\label{QD4}
Q_{D4}=\frac{1}{(2\pi)^3}\int F_{(4)}= \left\{\begin{array}{l l}
\vspace{1 mm}
\frac{N_c^{3/2}e^{2\Delta\phi}}{\sqrt{c}\pi}\bigg[\frac{\sqrt{g_0}}{24}  r^2-\frac{\left(g_0-1\right) \left(21 g_0+25\right) }{864 g_0^{3/2}}r^4...\bigg]&\quad r\approx0\\
\vspace{1 mm}
\frac{N_c^{3/2}}{\pi}\bigg[\frac{8 c x-16+c^2 c_{\gamma }+\frac{8}{3} r (16-6 c x)}{4
   \sqrt{3} c^2}e^{-2r}+...\bigg]&\quad r\approx\infty\\
\end{array}\right.
\end{equation}
where this quantity has been pulled back onto the 4-cycle defined by fixing $\psi_1$ and $\psi_2$ such that $\psi_1-\psi_2=(2k+1)\pi$ for integer $k$. These angles refer to the representation of the left invariant 1-forms in Eq.\ref{rep}.

The water muddies at this point. There is no suitable 4-form combination with which to define a quantised page charge for the D4 brane. However The D4 page charge dies off very quickly in the UV as $Q_{D4}\sim e^{-2r}$ compared to $c_0\sim Q_{D2}\sim e^{2r/3}$. In fact numerical plot of $Q_{D4}$ in Fig.\ref{charges} shows that it is always small compared to these quantities and after first reaching a maximum close to the IR it once more dies off again very quickly. 

\section{On the Field Theory}\label{OntheFieldTheory}
In section \ref{OnTheSupergravity} a new radial variable was introduced which was used to solve the BPS system of Appendix \ref{BPSEQ}.  This does not represents a different SuGra solution to that presented in \cite{Macpherson:2012zy} but rather gives the best coordinate system to describe the solution. A UV expansion for the (generalised) deformed Maldacena-Nastase solution that persists to all orders in $r$ was presented and some aspects of the SuGra solution explored.

Using what has been learnt in the previous sections the objective from here on is to learn some new information about the dual field theory. Neither \cite{Gaillard:2010gy} nor \cite{Macpherson:2012zy} made much of an attempt in this direction so the results of this section are new for both the Gaillard-Martelli solution and its generalisation.

\subsection{Operator Analysis}\label{operators}
It was shown in Section \ref{ctoinf} that the metric in the UV is asymptotically $AdS_4\times Y$, where $Y$ is a compact space. However we cannot have a CFT living on the boundary of the space as the dilaton tends to $\phi\sim -4r/3$ rather than a constant. None the less there is some hope that some insight may be gleamed from the AdS-CFT dictionary. 

For a full study of the operators in the dual field theory one should integrate the type-IIA action over the compact part of the background then perform a holographic renormalisation group analysis on the resulting 4-d theory in the spirit of \cite{Heemskerk:2010hk}. However the literature on this subject deals with duals to 3+1d theories and deriving equivalent results for 2+1d theories is outside the scope of this work. 

The work of \cite{Polchinski:2000uf} which presents a more ad-hoc method for extracting information about the operator content of a field theory from its dual gravity description, as long as the metric is asymptotically AdS.

When $\rho$ is the standard AdS radius, ie $\rho=e^{2r/3}$, an asymptotic solution which behaves like:
\begin{equation}
a_i \rho^{\Delta-3}+b_i \rho^{-\Delta}
\end{equation}
corresponds to a dimension $\Delta$ operator insertion into the Lagrangian of the field theory of the form:
\begin{equation}
\mathcal{L'}=\mathcal{L}+a_i\mathcal{O}_i
\end{equation}
A non zero $b_i$ then implies that this operator is picking up a vev:
\begin{equation}
<0|\mathcal{O}_i|0>=b_i
\end{equation}
The case $a_i=0$ with $b_i\neq0$ can also be interpreted as a condensate.

Thus it is the sub leading terms in the UV expansions of the various functions that make up the metric that this analysis will be performed on. That is the leading order non $AdS_4$ deformations. The relevant combinations, written in terms of $\rho=e^{2r/3}$ are:
\begin{equation}\label{oper}
\left.\begin{array}{l l}
\vspace{ 3mm}
\frac{1}{\sqrt{H}}&=\frac{\rho ^2 N_c}{\sqrt{3 c x+4}}+\frac{2 N_c (c x+2)}{c (3 c x+4)^{3/2}}+\frac{N_c (9 c x (c x (120 c
   x+101)-244)-1856)}{48 c^2 \rho ^2 (3 c x+4)^{5/2}}+O\left(\frac{1}{\rho }\right)^4\\
\vspace{ 3mm}
\sqrt{H}e^{2g}\frac{dr}{d\rho}^2&=\frac{9 N_c \sqrt{3 c x+4}}{4 \rho ^2}-\frac{9 \left(N_c (5 c x+8)\right)}{4 \rho ^4 \left(c \sqrt{3 c
   x+4}\right)}+O\left(\frac{1}{\rho }\right)^5\\
\vspace{ 3mm}
\frac{\sqrt{H}e^{2g}}{4}&=\frac{3}{16} N_c \sqrt{3 c x+4}+\frac{3 N_c (7 c x+8)}{16 c \rho ^2 \sqrt{3 c x+4}}+O\left(\frac{1}{\rho
   }\right)^4\\
\vspace{ 3mm}
\frac{\sqrt{H}e^{2h}}{4}&=\frac{1}{4} N_c \sqrt{3 c x+4}-\frac{N_c (5 c x+8)}{4 \rho ^2 \left(c \sqrt{3 c
   x+4}\right)}+O\left(\frac{1}{\rho }\right)^4\\
\end{array}\right.
\end{equation}
these are the factors of $dx_{1,2}^2$, $d\rho^2$, and the two 3-spheres in the metric respectively. The leading terms of the first and second equations are the $AdS$ terms so these will be ignored.

It is possible to see two different behaviours in Eq.\ref{oper}. The sub leading term of the first equation signals a $\Delta=3$ operator insertion in the Lagrangian of the theory. This operator is marginal and does not have a vev in the solutions we consider in the bulk of this work. However $\rho^{-3}$ terms that would give a vev to this operator appear in the general UV expansions of Appendix \ref{A2}. We see this behaviour once more in the leading terms of the last 2 equations. It is interesting to note that this behaviour is also seen in the UV expansion of $\gamma$ (see Eq.\ref{UVexpansions} and Eq.\ref{GenUVSer}). We also see a  $\Delta=4$ vev in the sub leading factor of $d\rho^2$, it possible that this controls the IR dynamics of the theory.

\subsection{Wilson Loops}\label{WilsonLoops}
In this section we will calculate the inter-quark potential between two massive non dynamical quarks. This can be extracted from the expectation value of a rectangular Wilson loop which extends in time $T$ and space $L$ in the $T\to\infty$ limit. The spacial extent of the loop defines the separation of a quark/anti-quark pair and the potential $E(L)$ can be found via the identification:
\begin{equation}
<\mathcal{W}_{\Box}>\sim e^{-T E(L)}
\end{equation}
Wilson loops are an effective tool for probing the behaviour of a gauge theory. In particular they will obey an area law in the IR if the theory exhibits confinement, where by it is only the area rather than the precise shape of the loop that will determine its expectation value. If a theory is conformal the expectation value of a loop has to have a particular form determined by conformal invariance. In terms of the inter quark potential this amounts to the following behaviours:
\begin{itemize}
\item Confining behaviour: $E(L)\sim L$
\item Conformal behaviour: $E(L)\sim\frac{1}{L}$
\end{itemize}

In the gauge-gravity correspondence, the gravitational dual of a Wilson loop is a minimal surface which extends from a D-brane in the UV down into the IR \cite{Maldacena:1998im}. The UV boundary of this surface is given by the shape of the loop in the field theory with larger loops corresponding to surfaces that extend further along the holographic coordinate, and thus probe deeper into the IR. When the minimal surface has sufficient symmetry to be effectively 2-d we need only consider the Nambu-Goto action of a fundamental string so that:
\begin{equation}
E(L)=\frac{1}{T}\mathcal{S}_{\text{N.G}}
\end{equation}
This string will be fixed in the UV at $r=\infty$ and extend down to some finite $r_{min}\geq0$. The string also has to satisfy boundary conditions in the UV such that the coordinates that are parallel $x_{||}$ and perpendicular $x_{\perp}$ to the brane satisfy $\frac{dx_{||}}{dx_{\perp}} = 0$.

The method of calculating a rectangular Wilson loop via the Nambu-Goto action is well known for arbitrary metrics. \cite{Nunez:2009da} presents a rigorous derivation and contains extensive discussion that will not be repeated here. In order to get the correct normalisation factor for the Wilson loops the field theory coordinates must be rescaled such that they no longer have $N_c$ as a factor. The relevant functions are then:
\begin{equation}\label{fg}
f(\rho)^2=g_{xx}g_{tt}=\frac{1}{c^2H};~~~~g(\rho)^2=g_{xx}g_{\rho\rho}= \frac{N_c e^{2g}}{c} ;~~~~ V= \frac{f(\rho)\sqrt{f(\rho)^2-f(\rho_{min})^2}}{f(\rho_{min}) g(\rho)}
\end{equation}
and the length and potential of the rectangular Wilson loop are given by:
\begin{equation}\label{LandE}
\begin{split}
L=&2\int_{\rho_{min}}^{\infty}\frac{d\rho}{V};\\
\vspace{3 mm}
E=& f(\rho_{min})L+2\int_{\rho_{min}}^{\infty}\frac{g(\rho)}{f(\rho)}(\sqrt{f(\rho)^2-f(\rho_{min})^2}-f(\rho))d\rho-2\int_0^{\rho_{min}}g(\rho) d\rho 
\end{split}
\end{equation}
where the expression for the potential has had the infinite quark mass's subtracted and been expressed in a way such that each term in the sum is finite. Notice that the UV parameter $c$ appears in Eq. \ref{fg}. This must be numerically determined for every combination of $(g_0,x,r_*)$ under consideration as described towards the end of Section \ref{expansions}. The strings boundary conditions at infinity amount to $V\to\infty$ as $\rho\to\infty$ which can easily be shown to be satisfied (This would not be the case for finite $\beta$).

In Fig.\ref{WilsonLoops} there are numerical plots of the inter-quark potential $E(L)$ for various values of $(g_0,x,r_*)$. A detailed numerical study suggests that for all finite values of these parameters we are led to inverse power law behaviour in the UV and linear behaviour in the IR which is consistent with the dual QFT exhibiting confinement. Further more, given $(g_0,x)$ there exists $r_*=r_{crit}$ such that when $r_*<r_{crit}$ the transition between UV and IR behaviours is completely smooth. However the existence of the intermediate scale can be made manifest when $r_*>r_{crit}$. Then a first order phase transition appears just like for the Gibbs free energy vs. pressure curve of the Van der Waals gas. This behaviour was first observed in \cite{Bigazzi:2008gd} (see also \cite{Warschawski:2012sx} for wilson loop calculations in a similar set up).

Given $(x,r_*)$ the effect of increasing $g_0$ is more subtle. When $x=0$ there is little effect at all, however for $x>0$ increasing $g_0$ decreases the gradient of the linear behaviour in the IR, a sign that the QCD-like string tension is decreasing in the dual QFT. Increasing x for fixed $(g_0,r_*)$ also decreases the gradient of the potential in the IR and so the sting tension reduces also. 
\begin{figure}
 \begin{center}             
\includegraphics[width=1\textwidth]{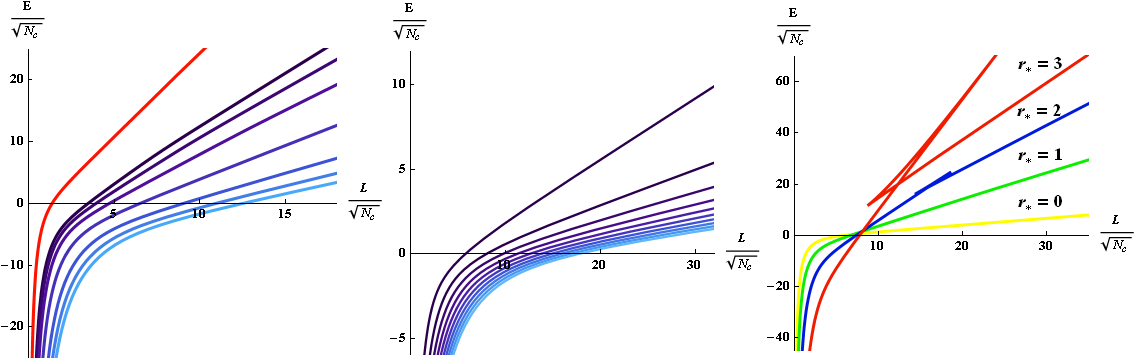}    
  \caption{Plots of potential verses Length for rectangular Wilson loops. The left panel shows plots for $g_0=10$, $r_*=1$ and $x=1/4,1/3,1/2,1,2,3,4$ colour coded purple to blue, $x=0$ is also shown in red. The middle panel has plots for $r_*=0$, $x=1$ and $g_0$ from 10 to 100 coded purple to blue once more. The right panel contains plots for $g_0=20$, $x=1/2$ for increasing $\rho_*$} 
  \label{WilsonLoops}
 \end{center}
\end{figure}

We can gain further insight into the asymptotic behaviours of the rectangular wilson loop using the expansions of section \ref{expansions}. The upper limit of the integral that defines L in Eq.\ref{LandE} is finite ($L\sim\int^{\infty}e^{-2r}dr$) while the lower limit is given by:
\begin{equation}\label{LIR}
L\sim\int_{\rho_{min}\sim0}(8 \sqrt{3 N_c} g_0 \sqrt{\frac{c g_0}{576 g_0 x+7}}\sinh (\Delta \phi )\frac{1}{r}+...)\sim \log (r)
\end{equation}
where we have have introduced $\Delta\phi = \phi_{\infty}-\phi_0$. Eq.\ref{LIR} diverges for small $r$ which indicates the absence of screening.

In \cite{Nunez:2009da} an exact expression for the rate of change of $E$ with respect to $L$ is derived:
\begin{equation}\label{dEdL}
\frac{dE}{dL}=f(\rho_{min})
\end{equation}
This equation can be used to derive an exact expression for the inter quark potential in terms of an expansion in large L, provided $r_{min}(L)$ can be found. In this case it can, we need only invert Eq.\ref{LIR} and integrate Eq.\ref{dEdL} to arrive at:
\begin{equation}\label{exactE}
E=\frac{1}{c \sqrt{1-e^{2 \Delta \phi }}}L+...
\end{equation} 
The next term is a complicated power of $e^{-L}$ that will not be quoted explicitly. When L becomes large, and so the deep IR of the field theory is being probed, this is a good approximation. Thus, in Eq.\ref{exactE}, we explicitly see that in the IR the potential is linear and so consistent with confinement with QCD-string tension, $\sigma$, given by:
\begin{equation}
\sigma=\frac{1}{c \sqrt{1-e^{2 \Delta \phi }}}
\end{equation}

At this stage we might want to ask about the $c\to\infty$ limit. We then have the exact solution described in Section \ref{ctoinf} and the relevant part of the metric is $AdS_4$:
\begin{equation}
f^2 = \left(\frac{1}{2}\right)^2 \rho^4;~~~~~g^2=N_c\left(\frac{9}{2}\right)^2 
\end{equation}
from this we can solve exactly for E, as in \cite{Maldacena:1998im}, and arrive at:
\begin{equation}\label{EAdS}
E=-(2\pi)^3\frac{81 \sqrt{N_c}}{2\Gamma(\frac{1}{4})^2L}
\end{equation}
Where $\rho=e^{2r/3}$. 

This result, although preordained by the virtues of $AdS$, is curious. It is a sign of conformal invariance in a theory that we know cannot be, because of dilaton scales as $\phi\sim-2r/3$. Of course it should be noted that the metric of the theory is only asymptotically $AdS_4$ in string frame and this is what the probe strings considered here see.

\subsection{(Gauge) Couplings}\label{couplings}
The generalised Gaillard-Martelli solution presented in Section \ref{OnTheSupergravity} have some striking similarities to the (Baryonic branch \cite{Butti:2004pk}) of Klebanov-Strassler \cite{Klebanov:2000hb} and its generalisation \cite{Conde:2011aa}. There is a running flux of $C_{(3)}$ at infinity reminiscent of the running $B_{(2)}$ of Klebanov-Strassler and there are many types of branes contained in the setup, albeit in type-IIA rather than type-IIB. For this reason one might anticipate that the field theory dual will be a confining 2-node quiver which, being 2+1d will also have a Chern-Simons term like the Maldacena-Nastase solution \cite{Maldacena:2001pb}. If this is true it should be possible to define 2 gauge couplings and this section will move towards that aim.

\begin{figure}
 \begin{center}             
\includegraphics[width=1\textwidth]{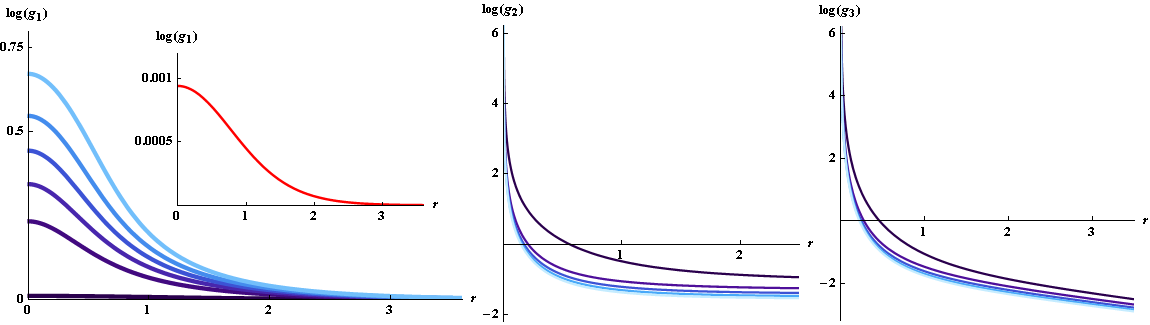}    
  \caption{Log plots of 3 putative gauge couplings. The left panel shows $Log(g_1)$ for $g_0=10$, $r_*=0$ with $x=0,1,2,3,4,5$ color coded purple to blue. As $x=0$ is very small it has been plotted again on a different scale in red. The centre and right panels shows $Log(g_2)$ and $Log(g_3)$ respectively for $g_0=20$, $r_*=0$ with $x$ increasing from 0 to 1/2 in increments of 1/8.} 
  \label{3couplings}
 \end{center}
\end{figure}

Consider a probe D2 brane which extends along the field theory directions with world volume flux $F$. The DBI action of such a brane is given by:
\begin{equation}
S_{D2}=T_{D2}\int d^3x\sqrt{-Det(\hat{G}_3+F)}
\end{equation}
if this is expanded to order $F^2$, the coefficient at that order defines a gauge coupling:
\begin{equation}
\frac{1}{g^2_1}= e^{-\phi}\sqrt{-Det(\hat{G}_3)}H = e^{\phi_\infty-\phi^{(0)}}
\end{equation}
where $\hat{G}_3$ is the pull back of the metric onto $(t,x,y)$, the factor H comes from $F_{\mu\nu}F^{\mu\nu}= H \eta^{\mu\alpha}\eta^{\nu\beta}F_{\mu\nu}F_{\alpha\beta}$ and we ignore constants factors. Clearly $\frac{1}{g^2_1}\sim1$ in the UV which is a sign of dilation invariance in this coupling, it cannot be full conformal invariance as the dilaton depends on $r$. Curiously the coupling is also constant in the IR, $\frac{1}{g^2_1}\sim e^{\Delta\phi}$, and the coupling smoothly interpolates between these values as shown in Fig.\ref{3couplings}.

Another coupling may be defined via a probe D4-brane extended along the field theory directions and wrapping the 2-cycle defined by $\theta_1=\theta_2$, $\phi_1=\phi_2$, $\psi_1=\psi_2=$constant. The relationship between these angles and the left invariant 1 form is given in Eq.\ref{rep}. This coupling is given by:
\begin{equation}
\frac{1}{g^2_2}= e^{-\phi}\sqrt{-Det(\hat{G}_5)}H = \sqrt{H}\left(4e^{2h}+e^{2g}(1-w)^2\right)e^{\phi_\infty-\phi^{(0)}}
\end{equation}
where constant factors are once more ignored and $\hat{G}_5$ is the induced metric. This coupling is also constant in the UV, $\frac{1}{g^2_2}\sim8\sqrt{4+3cx}$, however in the IR the behaviour is strongly coupled, $\frac{1}{g^2_2}\sim r^2$. See Fig.\ref{3couplings} for a numerical plot.

These couplings are sufficient for a 2-node quiver, however there is one further possibility. The Gaillard-Martelli solution also contains fractional D2-branes which come from NS5-branes wrapping the cycle $\sigma^i=\omega^i$ \cite{Macpherson:2012zy}. It is possible to define another coupling via a probe D2 instanton that wraps the 3-cycle on which the NS5-branes are wrapped\footnote{A D1 instanton is used to define the coupling of $\mathcal{N}=1$ SQCD in \cite{Casero:2006pt} however this is in type-IIB SuGra with a geometry generated from D5-branes which wrap a 2-cycle.}. This object must be dual to a Euclidean field theory object that is localised in both time and space, ie an instanton. Thus the identification $e^{S_{D2}}=e^{\frac{8\pi^2}{g^3_3} n}$, where n is the instanton number, is made. This coupling, up to constant factors, is given by:
\begin{equation}
\frac{1}{g_3^2}=e^{-\phi}\sqrt{-Det(\hat{G'}_3)}=\frac{\sqrt{H}e^{\phi_\infty-\phi^{(0)}}}{\sqrt{c}}\left(4e^{2h}+e^{2g}(1-w)^2\right)^{3/2}
\end{equation}
where $\hat{G'}_3$ is the induced metric on $\sigma^i=\omega^i$. This coupling is asymptotically free $\frac{1}{g_3^2}\sim e^{2r/3}$ as well as being strongly coupled in the IR where $\frac{1}{g_3^2}\sim r^3$ and is numerically plotted in Fig.\ref{3couplings}.

As shown in Fig.\ref{3couplings} $g_1$ and $g_2$ both asymptote to constants in the UV. This is inherited from the conformal symmetry of $AdS_4$, and while the symmetry is broken by the dilaton, this is evidence that the dual gauge theory has some residual dilation invariance at high energies. The coupling $g_3$ is different, it is asymptotically free. The fact that 3 couplings can be defined may at first sight seem strange since the claim is that the dual field theory is a 2 node quiver. The astute reader will notice however that none of them have been referred to as gauge couplings thus far, this is deliberate. 

At this stage it will be instructive to recall the form of the gauge coupling in Klebanov-Strassler. This is an $SU(N_l)\times SU(N_s)$ gauge theory with gauge couplings $g_l$ and $g_s$ defined as:
\begin{equation}
\begin{split}
\frac{1}{g^2_+}=\frac{1}{g^2_l}+\frac{1}{g^2_s}&\sim e^{-\phi^{KS}}\\
\frac{1}{g^2_-}=\frac{1}{g^2_l}-\frac{1}{g^2_s}&\sim e^{-\phi^{KS}}\int B^{KS}_2\\
\end{split} 
\end{equation}
The dual gravitational description is in terms of a geometry generated from D3 branes and fractional D3 branes\footnote{Actually this is done in the $\mathcal{N}=2$ orbifold case, and assumed to hold for the less supersymmetric example of Klebanov-Strassler}. These identifications can be made by studying the field theory living on these objects in a similar way to what is described above. The main point here is that $g_{+/-}$, the objects defined through a probe brane analysis like above, are not actually the gauge couplings. 

We expect a similar, albeit possibly more complex, scenario to be true for the (generalised) Gaillard-Martelli solution. We do not know enough about the dual QFT to be able to confirm certain proposals are correct, so one is limited to making more general comments. It is important to realise that the dual field theory can also contain Yukawa couplings as well as gauge couplings. These would be remnants from the higher dimensional origin of the fractional D2 branes in the background. The remainder of this section will now focus on what can be said about the putative couplings defined above.

In the IR the fact that $g_1$ starts to increase then becomes constant may appear strange, particularly in the light of the confining Wilson loops of Section \ref{WilsonLoops}. However one must appreciate that in YM-CS like theory's the coupling of the Yang-Mills term can increase as one flows toward the IR and then become frozen in the strong coupling regime. Here the dynamics will become dominated by a confining pure Chern-Simons theory. This mechanism is governed by the effective mass that the Chern-Simons level gives the gauge field, $g_{YM}^2|k|$. It is gratifying to see that the interpolating region of $g_1$ is of approximately the same width as that of $\gamma$ and $w$ in Fig.\ref{BPSNumerics}, this suggests that one of these functions is dual to an object in the gauge theory that governs this scale\footnote{A similar function controls the condensate in the dual to SQCD so this is a reasonable assumption}. Thus it seems likely that this coupling is indeed of a Maxwell type.

The IR behaviour of $g_2$ is exactly what one would expect from a theory exhibiting confinement, a good job considering the conclusion of Section \ref{WilsonLoops}. Over the whole range of $r$, $g_2$ has the best behaviour one can ever hope for in a coupling. Gravitational duals to theories that are strongly coupled in the IR are highly desirable for obvious phenomenological reasons. However, at least in $3+1d$, there are no fully controlled geometries dual to $\mathcal{N}=1$ theories that flow to a conformal fixed point in the UV and confine in the IR. This can happen here because the metric is asymptotically $AdS_4$ and the warp factors in the metric and dilaton can conspire to cancel the asymptotic $r$ dependence of $e^{2h}$ and $e^{2g}$. The most attractive interpretation is that $g_2$ has the full dynamics of the theory encoded into it, however this would need to be checked against a calculation in the dual QFT.

The last coupling, $g_3$, bucks the trend of dilation invariance in the UV, it is asymptotically free. Note that this does not mean that the theory is weakly coupled here, as $g_1$ and $g_2$ demonstrate, it is not. Indeed, if it were the full 10d geometry would be highly curved and the SuGra approximation would not be valid. Instead this is a coupling which must be sensitive to only part of the dynamics of the theory. When the full theory is considered this effect must be overtaken by something else. It is interesting to note that the combination $g_1+g_3\sim g_2$ in both the IR and UV which gives some heuristic motivation to this argument.

\section{Discussion}\label{Diss}
In this work a generalisation of the Gaillard-Martelli solution, \cite{Gaillard:2010gy} with an additional intermediate scale was considered which is a solution to type-IIA SuGra. This was first derived in \cite{Macpherson:2012zy} by applying a $G_2$ structure rotation (equivalent to a U-duality) to a generalisation of the deformed Maldacena-Nastase solution \cite{Maldacena:2001pb,Canoura:2008at} which includes a 5 brane profile.

An new radial coordinate was introduced that gives a UV series solution to the BPS equations of the (generalised) deformed Maldacena-Nastase that persists to all orders in exponentials and polynomials. This was shown to numerically match the IR expansion and it was shown that (at least for all cases considered) there was only 1 tunable constant in both the IR and the UV. This puts the $G_2$ cone solutions on a more equal footing with their Conifold cousins however full equality would require a partial integration which is still lacking. The metric of the solution is asymptotically $AdS_4\times Y$, where $Y$ is the compact metric at the base of a $G_2$ cone. Conformal symmetry is broken however as the dilaton depends on the radial coordinate as $\phi\sim-r$. In the limit where the UV parameter $c\to\infty$ the semi analytic UV series solutions become exact and the metric is precisely $AdS_4\times Y$ with dilaton $\phi=1/2(\log 2 -4r/3)$, which unlike the result of \cite{Gaillard:2010gy} is IR finite. The exact solution has Maxwell charges for D2 and NS5 branes with the D2 charge running such that it is proportional to the flux $C_{(3)}$ on a certain cycle. A Page charge is introduced for the D2 brane which is quantised and can only be defined up to shifts induced by large gauge transformations on $C_{(3)}$. The story is similar for the full generalisation of the Gaillard-Martelli solution however a Maxwell charge for D4 branes can now be defined in addition to D2 and NS5 branes. The Maxwell charge for D4 branes runs in such a way that it grows from the IR but then dies off again before the UV. Thus the D4 branes appear to be localised near the IR.

It is likely that the Gaillard-Martelli solution and its generalisation are dual to a 2 node quiver with gauge group $SU(N)\times SU(M)$. This is because of the similarity it bares to the baryonic baryonic branch of Klebanov-Strassler \cite{Klebanov:2000hb,Butti:2004pk}. Klebanov-Strassler also exhibits a duality cascade which could be mediated in the Gaillard-Martelli solution via the shifts in the D2 page charge under large gauge transformations, as in \cite{Benini:2007gx}. Proving this however is outside the scope of this work. 

An operator analysis was performed which suggests that the Lagrangian of the dual field theory contains a dimension 3 operator insertion. It also indicated that there was a dimension 4 vev, which may be partially responsible for the IR dynamics.

It was shown that the dual gauge theory is confining via a study of rectangular Wilson loops.  Unfortunately it is not clear how to back this result up with a calculation of the $k$-string tension in the spirit of \cite{Herzog:2001fq}. This is because in the IR it is possible to perform a gauge transformation such that the compact part of the metric is a round $S^3$ and we require an $S^3$ inside an $S^4$ to follow the prescription of \cite{Herzog:2001fq}. The Wilson loop study also signals the presence of the intermediate scale, which manifests itself through a first order phase transition in a certain region of parameter-space. The recent results of \cite{Warschawski:2012sx} indicate that it should be possible to add further intermediate scales by modifying the NS5-brane profile. This may be of some use in holographic condensed matter model building.

Further evidence of confinement was shown by the gauge coupling. This behaviour was interpreted as coming from a pure Chern-Simons theory which dominates the IR, the evidence for which is the freezing of a Maxwell like coupling there. Two more couplings were also defined which are both strongly coupled in the IR. The analysis suggests that the full theory exhibits some residual dilation symmetry at high energies as indicated in 2 of the couplings. However care should be take with these statements. It is not entirely clear exactly what combinations of these objects map to the gauge couplings in the dual QFT.

This work and \cite{Macpherson:2012zy} show that it is possible to derive solutions to the BPS equations of deformed Maldacena-Nastase in terms of 2 different radial variables. It would be desirable to find the asymptotically linear dilaton solutions of \cite{Macpherson:2012zy} in the radial variable used here. Perhaps this would lead to a greater understanding of that system.

An interesting future direction would be to try to find new IR solutions that can be numerically matched to the general UV expansions in the appendix. These solutions would still have metrics that are asymptotically $AdS_4$ for arbitrary values of $(c_h,c_w,c_\gamma)$. The question is which combinations of these values lead to a regular IR. Experience of similar Conifold solutions \cite{Nunez:2008wi,Nunez:2008wi} indicates that it should be possible find $G_2$ solutions which exhibit a similar walking type behaviour. A brief numerical investigation, solving from the UV, seems to suggest that this is so but a more detailed study is required and the IR must actually be derived.

It would also be of interest to calculate the scalar (glue ball) spectrum of this theory as this is one of the few examples of solutions that are both confining and asymptotically $AdS$, albeit only in string frame and in $2+1d$. An algorithm for doing this is presented in \cite{Elander:2010wd} and references therein which should be applicable, modulo small modifications accounting for the fact that the field theory dual here is not $3+1d$.

\section{Acknowledgements} 
I would like to thank Adi Armoni and Carlos N\'u\~nez for comments and useful discussions. I would also like to thank Ed Bennett and Daniel Schofield for comments.

\appendix
\section{The BPS equations}\label{BPSEQ}
Here we quote the BPS equations for the definitions of the fields we use here. These results were derived for the flavour profile dependent system in \cite{Macpherson:2012zy} and are generalisations of the BPS systems derived in \cite{Gaillard:2010gy,Maldacena:2001pb,Canoura:2008at}. 

The $\beta\to0$ limit of the system we describe in section \ref{setup} gives the metric:
\begin{equation}
ds^2_{\text{str}} = N_c\bigg(\frac{1}{c}dx^2_{1,2}+e^{2g}dr^2+\frac{e^{2h}}{4}(\sigma^i)^2+\frac{e^{2g}}{4}(\omega^i-\frac{1}{2}(1+w)\sigma^i)^2\bigg)
\end{equation}
with $H_{(3)}$ defined as in Eq.\ref{H3F4} and $F_{(4)}=0$. The conditions that  $\mathcal{N}=1$ supersymmetry is satisfied in this limit imply the following differential equations for the various functions in the metric and $H_{(3)}$:
\begin{equation}\label{bps}
\left.\begin{array}{l l} 
\vspace{3 mm}
\phi'^{(0)} &=\frac{1}{8} \bigg[-V e^{g-3 h} \sin (\alpha )+12 e^{-g-h} (\gamma -w) \sin (\alpha )+8 e^{-2 g} \cos
   (\alpha )+\\
   \vspace{3 mm}
  & ~~~~~~6 e^{-2 h} \cos (\alpha ) \left(4 P x-w^2+2 w \gamma -1\right)\bigg]\\
\vspace{3 mm}             
h'&=\frac{1}{8} e^{-g-3 h} \bigg[-4 e^{g+h} \cos (\alpha ) \left(e^{2 g} \left(w^2-1\right)+4 P x-w^2+2 w
   \gamma -1\right)-\\
   \vspace{4 mm}     
  & ~~~~~~4 e^{2 h} \sin (\alpha ) \left(\left(2 e^{2 g}-1\right) w+\gamma \right)+e^{2 g} V
   \sin (\alpha )\bigg]\\   
   \vspace{4 mm}
g'&=   \frac{1}{4} e^{-2 h} \cos (\alpha ) \left(e^{2 g} \left(w^2-1\right)-4 P x+w^2-2 w \gamma
   +1\right)+\\
   \vspace{3 mm}
   \vspace{2 mm}     
   &~~~~~~e^{-g-h} (w-\gamma ) \sin (\alpha )+\left(1-e^{-2 g}\right) \cos (\alpha )\\
\vspace{3 mm}             
 w'&=\frac{1}{4} \Bigg[(2 e^{-g-h} \sin (\alpha) \left(3 e^{2 g}
   \left(w^2-1\right)+4xP+2 w \gamma -w^2-1\right)+\\
   \vspace{3 mm}
   &~~~~~~8 \left(e^{2 g}-1\right)
   e^{h-3 g} \sin (\alpha )+4 \cos (\alpha ) \left(e^{-2 g} (\gamma -w)-2
   w\right)+e^{-2 h} V \cos (\alpha )\bigg]\\     
   \vspace{3 mm}
\gamma'&=-\left(w^2-1\right) e^{3 g-h} \sin (\alpha)+4 e^{g+h} \sin (\alpha)+4 e^{2 g} w \cos (\alpha
   )-4 x \eta  P'
\end{array}\right.
\end{equation}
Where $'$ refers to differentiation with respect to $r$. The trigonometric functions are defined through the following relation:
\begin{equation}\label{alpha}
\tan(\alpha) =\frac{V e^{3 g-h}-12 e^{g+h} \left(\left(2 e^{2 g}-1\right) w+\gamma \right)}{6 e^{2 g} \left(4 e^{2 h}-4
   P x+w^2-2 w \gamma +1\right)-6 e^{4 g} \left(w^2-1\right)-8 e^{2 h}}
\end{equation}
And the functions $V$ and $\eta$ are defined as:
\begin{equation}\label{VAndeta}
\left.\begin{array}{l l} 
\vspace{3 mm}
\eta &= \frac{e^g (C+w)+2 e^h \tan (\alpha )}{-4 e^{2 h-g}+e^g \left(w^2-1\right)+4 e^h w \tan (\alpha )}\\
V&=(1-w^2)(w-3\gamma)-4(1-3xP)w+8(\kappa+\frac{3x}{2}C)
\end{array}\right.
\end{equation}
In the main part of this work we have set:
\begin{equation}
C=1
\end{equation}
As it seems that this is required to have a quantised Chern-Simons term when $P$ depends on $r$ and we also choose:
\begin{equation}
\kappa=\frac{1}{2}
\end{equation}
To avoid a curvature singularity in the IR.

It is possible to show that if we can solve the BPS equations at $\beta=0$ then this solution will automatically solve the more complicated BPS equation of the system described in section \ref{setup} where $\beta$ is arbitrary (see \cite{Gaillard:2010gy} for details).

\section{The General Asymptotic UV Solution to the BPS Equation}\label{A2}
In the UV, the general series solution to the BPS system of Appendix \ref{BPSEQ} with $C=1$ and $\kappa=\frac{1}{2}$ is given by:
\begin{equation}\label{GenUVSer}
\left.\begin{array}{l l} 
   \vspace{3 mm}
e^{2g}&=c \left(e^{2 r/3}\right)^2-1-\frac{2 c_h}{3 e^{2 r/3}}+\frac{33}{4 c \left(e^{2 r/3}\right)^2}+\frac{\frac{27
   c_h}{5 c}-\frac{8 c_w}{15}}{\left(e^{2 r/3}\right)^3}+\frac{\frac{8 c c_h^2+49 c x-396}{9 c^2}-\frac{c
   c_w^2}{24}}{\left(e^{2 r/3}\right)^4}+\\
      \vspace{3 mm}
   &~~~~\frac{3 c_h (20 c x-537)+160 c c_w}{42 c^2 \left(e^{2
   r/3}\right)^5}+\\
      \vspace{3 mm}
   &~~~\frac{15 c \left(864 c c_h c_w-6048 c_h^2+c \left(285 c c_w^2-56 c_{\gamma }\right)\right)+4
   (1120 r (3 c x-8)+3 c x (495 c x-32734)+465192)}{7200 c^3 \left(e^{2 r/3}\right)^6}+\\
      \vspace{3 mm}
   &~~~\frac{c_h \left(765 c^3
   c_w^2-105440 c x+943758\right)-4480 c c_h^3-16 c c_w (635 c x+1309)}{3240 c^3 \left(e^{2
   r/3}\right)^7}+...\\
   \vspace{3 mm}
e^{2h}&=\frac{3}{4} c \left(e^{2 r/3}\right)^2+\frac{9}{4}+\frac{c_h}{e^{2 r/3}}+\frac{3 x-\frac{77}{16 c}}{\left(e^{2
   r/3}\right)^2}+\frac{3 \left(2 c_w-\frac{9 c_h}{c}\right)}{10 \left(e^{2 r/3}\right)^3}-\frac{3 c^3 c_w^2+32 c
   c_h^2+88 c x-1536}{96 c^2 \left(e^{2 r/3}\right)^4}+\\
      \vspace{3 mm}
   &~~~\frac{c_h (863-36 c x)+6 c c_w (21 c x-79)}{84 c^2
   \left(e^{2 r/3}\right)^5}+\\
      \vspace{3 mm}
   &~~~\frac{15 c \left(-416 c c_h c_w+672 c_h^2+3 c \left(8 c_{\gamma }-75 c
   c_w^2\right)\right)-4 (480 r (3 c x-8)+3 c x (1755 c x-14006)+99728)}{3200 c^3 \left(e^{2
   r/3}\right)^6}+\\
      \vspace{3 mm}
   &~~~\frac{c_h \left(-765 c^3 c_w^2+37280 c x-206262\right)+640 c c_h^3+4 c c_w (5165 c
   x-15326)}{2160 c^3 \left(e^{2 r/3}\right)^7}+...\\
   \vspace{3 mm}
w&=\frac{2}{c \left(e^{2 r/3}\right)^2}+\frac{c_w}{\left(e^{2 r/3}\right)^3}+\frac{22-6 c x}{c^2 \left(e^{2
   r/3}\right)^4}+\frac{32 c_h+27 c c_w}{6 c^2 \left(e^{2 r/3}\right)^5}+\frac{3 c^2 c_h c_w-16 c x+51}{2 c^3
   \left(e^{2 r/3}\right)^6}+\\
      \vspace{3 mm}
  &~~~ \frac{c_h (436-165 c x)+c c_w (45 c x+7)}{30 c^3 \left(e^{2r/3}\right)^7}+...\\
   \vspace{3 mm}
\gamma&=\frac{1}{3}+\frac{x}{\left(e^{2 r/3}\right)^2}+\frac{2 c_w-\frac{2 c_h}{c}}{\left(e^{2
   r/3}\right)^3}+\frac{\frac{16 r (8-3 c x)}{3 c^2}+c_{\gamma }}{\left(e^{2 r/3}\right)^4}+\frac{3 c_h (c
   x-7)+c c_w (c x-11)}{c^2 \left(e^{2 r/3}\right)^5}+\\
      \vspace{3 mm}
   &~~~\frac{c^2 \left(-9 c_h c_w+c c_w^2-6 c_{\gamma }\right)+4
   (8 r (3 c x-8)-7 c x+24)}{3 c^3 \left(e^{2 r/3}\right)^6}+\\
      \vspace{3 mm}
  &~~~ \frac{c_h \left(-12 c^2 c_{\gamma }+64 r (3 c
   x-8)-207 c x+930\right)+9 c c_w (78-5 c x)}{18 c^3 \left(e^{2 r/3}\right)^7}+...\\
      \vspace{3 mm}
   e^{2(\phi-\phi_\infty)}&=1+\frac{3 c x+4}{c^2 \left(e^{2 r/3}\right)^4}+\frac{2 \left(c_w+3\right)}{5 c \left(e^{2 r/3}\right)^5}-\frac{4
   (c x+2)}{c^3 \left(e^{2 r/3}\right)^6}+\frac{3 c_w (6 c x+1)-48 c x-143}{21 c^2 \left(e^{2
   r/3}\right)^7}+\\
         \vspace{3 mm}
   &~~~\frac{-8 c^3 c_w-3 c \left(8 c \left(c+6 x^2\right)-283 x\right)+1136}{24 c^4 \left(e^{2
   r/3}\right)^8}+\frac{c_w (63-68 c x)+1546 c x+3429}{90 c^3 \left(e^{2 r/3}\right)^9}+\\
         \vspace{3 mm}
   &~~~\frac{3 c^3 c_w
   \left(c_w (15 c x-71)-60 c x+244\right)-20 \left(78 c^3+1267\right) c x+2468 c^3+5190 c^2 x^2-43050}{225 c^5
   \left(e^{2 r/3}\right)^{10}}+...
     \end{array}\right.
\end{equation}
As one would expect of a general solution of 4 coupled 1st order ODEs, these series depend on 4 integration constants $c$, $c_h$, $c_w$, $c_\gamma$. There is also $\phi_\infty$ but this is not independent. In the main part of this paper the choice $c_h=c_w=0$ was made, this sets all odd powers of $e^{2r/3}$ in $e^{2g}$, $e^{2h}$, $w$ and $\gamma$ to zero, simplifying the expansions considerably. When numerical matching of the IR and UV asymptotic solutions was performed it was also found that $c_\gamma=0$ in all cases considered.  It seems likely that solutions with more constants turned on do not match to the IR of Eq.\ref{IRExpansion1}.

Notice that the leading order of all the functions in Eq.\ref{GenUVSer} depend only on $c$ and so we must still have a metric that is asymptotically $AdS_4$ for arbitrary values of $(c_h,c_w,c_\gamma)$. The more relevant question is which combinations of these values lead to a regular IR. Experience of similar Conifold solutions \cite{Nunez:2008wi,Elander:2011mh} suggests it should be possible find $G_2$ cone solutions which exhibit a similar walking type behaviour. Confirming this however, shall be left to future work.


\end{document}